# Develop-Fair Use for Artificial Intelligence:

## A Sino-U.S. Copyright Law Comparison Based on the *Ultraman, Bartz v. Anthropic*, and *Kadrey v. Meta* Cases

Chanhou Lou[1]

**Abstract:** The static doctrine of fair use in traditional copyright law can no longer effectively respond to the challenges posed by generative artificial intelligence (AI). Drawing on a deep comparative analysis of China's *Ultraman* case and the U.S. *Bartz v. Anthropic* and *Kadrey v. Meta* cases, this article proposes the theory of "Develop-Fair Use" (DFU). This theory posits that fair use for AI (AIFU) is not a fixed category of exceptions but a dynamic tool of judicial balancing, whose analytical logic should shift from examining closed scenarios to applying an evaluative rule in open-ended contexts. The judicial focus, in turn, shifts from the formal subsumption of facts to the substantive balancing of competition in relevant markets. While the judicial practices in China and the U.S. follow different paths, they both reveal this inherent logic: the *Ultraman* case, through its judicial creation of a "four-context analysis," carves out institutional space for the AI industry's development; meanwhile, the debate in the two U.S. cases over the fourth factor, "market impact," particularly the "market dilution theory" proposed in *Kadrey*, broadens the scope of judicial review from direct substitution in copyright markets to the wider dimension of industrial competition. The core of the DFU paradigm lies in recognizing and balancing the "antinomy" of competition in relevant markets: namely, the dynamic interplay between the emerging AI industry, which asserts fair use to develop its relevant markets, and the traditional publishing industry, which cultivates its relevant markets, such as the one for "training licenses," to oppose fair use. Consequently, the boundary of fair use is not the result of pure legal deduction, but a case-specific factual determination based on a judicial assessment of the evolving realities of market competition. Ultimately, this approach aims to trim the excesses of copyright scope while remedying the insufficiencies of market competition.

---

[1] Dr. Chanhou Lou, Ph.D. in Law, Macau Fellow Researcher at the University of Macau, DLI Fellow Researcher at Cornell Tech, Cornell University, chlou@um.edu.mo, https://orcid.org/0009-0003-3494-1079, Working paper (preprint), v21, 8 Sep 2025 ©.

# Table of Contents

## I. Copyrightable Fair Use for AI?

Fair use trims copyright's overreach and fills deficits in market competition. It is not a fixed carve-out but what the author terms "Develop-Fair Use" (DFU). From a developmental vantage, the substantive value of copyright equals the residual remaining after subtracting the domain of fair use from the legal scope of protection. If an expansive reading of originality brings "artificial intelligence generated content" (AIGC) within the category of protected "works,"[i] a commensurately capacious account of fair-use scenarios for "generative artificial intelligence" (AI) must follow so that copyright's residual value remains in dynamic equilibrium. Fair use for AI (AIFU) is therefore linked to the growth of the AI sector. On 26 August 2025, the State Council of the People's Republic of China (P.R.C.) promulgated the national policy titled "Opinions on Deeply Implementing the 'AI+' Action," which centered the policy discourse of "development," calling for "strengthening data-supply innovation… improving data-property and copyright rules adapted to AI development, and lawfully opening copyright content created with public funds."[ii] This framing indicates why fair use under China's *Copyright Law* matters for AI:[iii] copyrightable works supply the "high-quality" training data AI needs. The upshot is that AI fair use is not static; it is the DFU signaled by rules "adapted to AI development."

By way of a comparison, the United State of America (U.S.) President Donald Trump's remarks on 23 July 2025 about "America's AI Action Plan" implicitly linked fair use to competitive capacity. He argued that: "You can't be expected to have a successful AI program when every single article, book, or anything else that you've read or studied, you're supposed to pay for…. [W]e have to allow AI to use that pool of knowledge without going through the complexity of contract negotiations," and he added that "China is not doing it," urging the U.S. to avoid self-imposed institutional limits.[iv] The premise is that a broader scope of AI fair use tends to strengthen competition in AI, which shifts the traditional fair-use balance between copyright and the public interest from an infringement setting to one centered on AI development and market competition. Policy statements alone cannot fix the boundaries of AIFU; those contours must be specified case by case in adjudication. As the Congressional Research Service advises: "Alternatively, given the limited time courts have had to address these issues, Congress may adopt a wait-and-see approach. As courts decide more cases involving generative AI, they may be able to provide greater guidance and predictability in this area."[v] Methodologically, AIFU is examined through a comparative analysis of P.R.C. and U.S. judicial decisions.

Academic writing on AIFU is abundant and generally falls into a three-part taxonomy. First is the constitution theory, which narrows the constitutive elements by characterizing uses as "non-expressive" or "non-work," thereby removing AI uses from the reach of copyright regulation.[vi] Second is the exception theory, which borrows the U.S. fair-use "four-factor test" or the comparative-law statutory exception for "Text and Data Mining" (TDM) to negate the illegality of consent-less AI uses.[vii] Third is the responsibility theory, which invokes notice-and-takedown rules like "Safe Harbor," contributory-infringement principles, and platform "duty of care" under network-tort liability to provide defenses for AI uses.[viii] Much of this literature is legislative rather than judicial and normative rather than empirical: it proposes ideal models of what AIFU should be, selectively tailoring and citing real-world scenarios to vindicate each position, while giving insufficient attention to the overall landscape of how AIFU operates in China and offering too little analysis of Chinese case law.

It is understandable that "even a smart chef cannot cook without rice": the paucity of case studies reflects the nascency of AIFU issues and the resulting scarcity of judicial materials. Using "artificial intelligence" (*rengong zhineng*) and "fair use" (*heli shiyong*) as keywords for an automated search in the "*Beida Fabao*" legal database,[ix] and then manually excluding noise such as "fair use fees" (*heli shiyong fei*), which is "reasonable fees for using copyrightable work," and non-copyright causes of action, yields very few published judgments. Many decisions, such as the Shanghai "AI face-swapping case," mention "fair use" only in passing, without analysis.[x] The sole decision with sustained reasoning is the Hangzhou *Ultraman* case, which did not reach a final judgment until December 30, 2024.[xi] A related matter that is the Guangzhou *Ultraman* case, billed as the "world's first generative AI service copyright infringement case," does not discuss fair use in its full-text judgment.[xii] Even so, this limited corpus does not prevent scholars from using the Hangzhou Ultraman texts and associated discourse as empirical entry points to reconstruct the overall landscape of AIFU latent in the deep structure of positive law and its specific contexts. Regrettably, existing studies of the Hangzhou *Ultraman* case have yet to provide such a thick description.[xiii]

In contrast, U.S. scholarship likewise confronts a paucity of judicial materials because courts are still addressing this emerging set of AIFU issues.[xiv] Pamela Samuelson undertook a prospective study of AIFU, offering a historical account of U.S. case law at the intersection of disruptive technological innovation and fair use.[xv] When her article appeared in 2024, however, the AIFU-related decisions in *Bartz v. Anthropic* and *Kadrey v. Meta* had not yet reached "summary judgment" in June 2025,[xvi] so her treatment of those two cases was necessarily predictive. Although Samuelson later released a short commentary "without citation notes" after the summary judgments, it did not advance a fully argued scholarly position; moreover, she expressed skepticism toward the "market dilution theory" developed in *Kadrey* and did not register the DFU implications of that theory arising from the expansion of copyright boundaries by AIGC. [xvii] Consequently, on both sides of the Pacific there remains no thick, empirical depiction of *Ultraman*, *Bartz*, and *Kadrey* cases; the current picture of how Chinese and U.S. courts are applying AIFU is still indistinct, and comparative analysis between the two jurisdictions is largely absent.

To address these gaps, this paper tests the proposition that "AIFU is DFU" through a comparative analysis of Sino–U.S. cases. The legal propositions of DFU are "interlinked": first, fair use does not operate through closed scenarios but within open-ended contexts; second, because the contexts are open, judicial focus shifts from threshold scenario testing to evaluative assessment, with particular attention to market-competition review; third, that review is not static, and requires defining relevant markets and balancing the antinomy between industries in that the emerging AI industry invoking fair use to develop relevant markets, and the traditional copyright industry developing relevant markets to oppose fair use.

In summary, the argument unfolds in three steps. First, it offers a thick description of China's AIFU practice through the Hangzhou *Ultraman* case. Second, it places that case alongside the U.S. *Bartz* and *Kadrey* cases to identify where the two systems converge and where they diverge in applying AIFU contexts. Third, it draws on the comparison to contend that AIFU is not a fixed exception, but DFU that adjusts over time with legal and market conditions.

## II. The *Ultraman* Case and the Chinese Landscape

The fair-use question in the Sino-Hangzhou *Ultraman* case is whether a platform's allowing users to train custom models and to publish those models and their outputs constitutes fair use, so that no copyright is infringed.[xviii] The plaintiff, Shanghai Cultural Development Co., Ltd. ("Cultural Company"), holds exclusive authorization for the copyrights in the Japanese anime superhero character images of the "Ultraman" series; and the defendant, Hangzhou Intelligent Technology Co., Ltd. ("Technology Company"), operates a generative-AI platform that lets users create the disputed "Ultraman"-related images in three ways.[xix]

First, model training. Users upload Ultraman-related images and, after adjusting base-model parameters, train a Low-Rank Adaptation ("LoRA") model whose style approximates that of the Ultraman works; the resulting model can be saved, applied, published, or shared by link.[xx] Notably, a LoRA model functions like a "small plug-in" attached to a large model: by modifying only a small set of parameters, it learns the Ultraman style, so that subsequent content generated by the LoRA model consistently maintains that style.

Second, model application. By entering Ultraman-related prompts, users select a base model or overlay a LoRA tuned to the Ultraman style; adjusting parameters then yields images like the Ultraman works, and the resulting images can be saved, published, downloaded, or shared via link.[xxi]

Third, mixed instructions. Users upload Ultraman-related images together with prompts to adjust base-model parameters; this produces images in the Ultraman style, which may likewise be saved, published, downloaded, or shared by link.[xxii]

The dispute unfolded as follows. The plaintiff Cultural Company alleged that the defendant Technology Company infringed the information network dissemination right in the Ultraman works and engaged in unfair competition and therefore filed suit with the Hangzhou Internet Court on February 20, 2024.[xxiii] On September 25, 2024, the court issued a first-instance judgment holding the defendant liable for copyright infringement; having afforded relief under the *Copyright Law*,[xxiv] it declined to reach the *Anti-Unfair Competition Law* claim.[xxv] Although the Cultural Company prevailed below, it appealed to the Zhejiang Province Hangzhou Intermediate People's Court ("Hangzhou Intermediate Court").[xxvi] On December 30, 2024, the Hangzhou Intermediate Court dismissed the appeal and affirmed the judgment.[xxvii]

Overall, although the first-instance court found copyright infringement, it also set out a fair-use analysis and, by disaggregating use contexts, concluded that AI data-training and content-generation contexts may constitute fair use. As the Hangzhou Internet Court stated: "In the absence of evidence that generative AI is intended to use the original expression of the copyrighted work, has affected the normal use of the work, or has unreasonably prejudiced the legitimate rights and interests of the relevant copyright owners, it may be deemed fair use."[xxviii] From this, the Chinese landscape of AIFU becomes apparent.

**2.1 Implicit Application of the Three-Step Test and Interpretive Dilemma**

The first-instance court implicitly applied the *Copyright Law*'s fair-use "three-step test" to ground AIFU….

### 2.2 Active Determination of Fair Use and Four-Context Analysis

First, the *Ultraman* first-instance court treated fair use not as an affirmative defense with the burden of proof on the defendant, but as a scope-of-rights question that the court examines *ex officio,* which is, by determining whether fair use applies, the court inversely delineates the boundaries of copyright.[xxix] Corroborating this approach, the defendant Technology Company did not invoke the fair-use clause in Article 24 of the *Copyright Law*, yet the court affirmatively found that the data-training context constituted fair use.[xxx] This active determination is linked to the court's recognition of the connection between non-expressive use and fair use, because the defendant implicitly advanced a non-expressive-use defense….

### 2.3 Does Fair Use Exclude Anti-Unfair Competition Law?

Compared with the first-instance judgment, the second-instance decision avoided a substantive discussion of fair use and addressed only a procedural point: "whether platform users of the Hangzhou Tech Company, without authorization from the rightsholder, use Ultraman works constitutes fair use is within the evaluation scope of the *Copyright Law*. In this appeal the Shanghai Culture Company expressly appeals only on the ground of unfair competition, and the actor in question is a platform user rather than the Hangzhou Tech Company, so this court will not comment further."[xxxi] It follows that the Hangzhou Intermediate Court confined the fair-use analysis to the user data-input context and, because users were not defendants in *Ultraman*, declined to evaluate whether that user data-input context constituted fair use….

## III. Comparison of Sino-U.S. Cases and Restatement of Develop-Fair Use

Like the Sino-Hangzhou *Ultraman* case, the U.S. cases *Bartz v. Anthropic* and *Kadrey v. Meta*, decided on 23 and 25 June 2025, respectively, also examined AIFU….

### 3.1 From Closed Scenarios to Open Contexts

Both the *Ultraman* and *Bartz* decisions adopt a contextual approach that distinguishes data input from data training….

### 3.2 From Scenario Rules to Evaluative Rules

The *Ultraman* case shows that article 24 of the *Copyright Law*, which states "and such use shall not affect the normal use of the work, nor unreasonably harm the copyright owner's lawful rights and interests," in the three-step test's evaluative rule does not preclude the four-factor test.[xxxii] The second step, "use of the work," corresponds to factor one, "the purpose and character of the use"; and the third step, "lawful rights and interests," corresponds to factor four, "the effect of the use upon the potential market for or value of the copyrighted work."[xxxiii] Compared with *Bartz*, *Kadrey* places greater emphasis on factor four, following the concurring opinion of Justice Neil M. Gorsuch in the U.S. Supreme Court's 2023 decision in *Andy Warhol Foundation v. Goldsmith* (hereinafter referred to as *Warhol*).[xxxiv] Although *Warhol* concerned only factor one, Justice Gorsuch looked back to the 1985 *Harper & Row Publishers v. Nation Enterprises* case, which said that "this last factor [factor four] is undoubtedly the single most important element of fair use,"[xxxv] and imported that view into factor one….

### 3.3 From Antinomy to Market Competition

The antinomy between fair use and market competition means this: the emerging AI industry invokes fair use to develop relevant markets, while the traditional copyright industry develops relevant markets to oppose fair use….

## IV. Conclusion

In sum, fair use for AI (AIFU) is not a fixed category of rights limitations, but what I call develop-fair use (DFU). The review logic shifts from testing closed scenarios to applying an evaluative rule in open-ended contexts, and the adjudicative focus moves from formal subsumption to substantive balancing of competition in relevant markets. Practice in China and the United States indicates this: China's *Ultraman* case, by extending a "four-context analysis," strategically preserves institutional space for AI-industry development; meanwhile, the divergence between *Bartz* and *Kadrey*, especially with *Kadrey*'s "market dilution theory," expands judicial review from directly substitutive copyright markets to the broader plane of industrial competition, together revealing a dynamic approach that moves beyond traditional copyright frames and faces market realities.

The core of DFU is to recognize and balance the antinomy of competition in relevant markets: a dynamic game between an emerging AI industry that invokes fair use to develop relevant markets and a traditional copyright industry that resists fair use by cultivating new arenas such as the training-license market. This requires adjudication to move beyond deductive copyright analysis and to incorporate unfair-competition analysis of evolving markets. Through such dynamic judicial balancing, fair use can function as an institutional tool that accommodates technological change and sectoral interests, restraining overreach in the scope of copyright while shoring up deficits in market competition, thereby serving the fundamental aims of innovation and knowledge dissemination.

In this respect, China and the United States present similar landscapes of AIFU. Neither jurisdiction hurried to give the AI industry a green light through statutory TDM exceptions;[xxxvi] instead, both legislatively "wait and see" by developing fair-use jurisprudence case by case, gradually feeling out the boundaries of industrial competition. For example, the proposed class settlement in *Bartz* illustrates DFU's explanatory power in two respects.[xxxvii] First, the USD 1.5 billion total with a uniform per-work price of USD 3,000 provides a concrete price signal and a measurement baseline for a training-license market.[xxxviii] The figure is comparable to the RMB 30,000 (approximately USD 4,200) copyright award in *Ultraman*, which supports empirical assessment of market impact and competition. Second, under a four-context analysis, the settlement resolves only input-side disputes through August 25, 2025, and expressly preserves output-side claims.[xxxix] This design channels the legality of content output and content use into later judicial review of market impact, aligns with DFU's shift from closed scenarios to open evaluation, and leaves room to balance post-settlement substitution and dilution effects.

Furthermore, this approach can flexibly accommodate, and iteratively adjust to, the effects of AI policy measures such as China's 2025 *Provisions on the Labelling of AI-Generated Synthetic Content* on the generated-dilution market and the training-license market, a topic reserved for future work.[xl] In short, DFU is not a fixed legal exception but a dynamic tool of judicial balancing, whose boundaries are shaped by the ebb and flow of market competition between the AI industry and the copyright industry, domestically and internationally. AI competition is therefore a competition over industries and over rules.